\documentclass[%
 aip,jcp,
 amsmath,amssymb,
 reprint,
]{revtex4-2}

\usepackage[a-1b]{pdfx} 

\usepackage{graphicx}% Include figure files
\usepackage{dcolumn}% Align table columns on decimal point
\usepackage{bm}% bold math
\usepackage[utf8]{inputenc}
\usepackage[T1]{fontenc}
\usepackage{mathptmx}
\usepackage{listings}
\usepackage{rotating} % Rotating table

\definecolor{codegreen}{rgb}{0,0.6,0}
\definecolor{codegray}{rgb}{0.5,0.5,0.5}
\definecolor{codepurple}{rgb}{0.58,0,0.82}
\definecolor{backcolour}{rgb}{0.95,0.95,0.92}
\lstdefinestyle{mystyle}{
  backgroundcolor=\color{backcolour},   commentstyle=\color{codegreen},
  keywordstyle=\color{magenta},
  numberstyle=\tiny\color{codegray},
  stringstyle=\color{codepurple},
  basicstyle=\ttfamily\scriptsize,
  breakatwhitespace=false,
  breaklines=true,
  captionpos=b,
  keepspaces=true,
  showspaces=false,
  showstringspaces=false,
  showtabs=false,
  xleftmargin=0.01\textwidth,
  rulecolor=\color[RGB]{200,200,200},
  frame=bt,
  framextopmargin=2pt,
  framexbottommargin=2pt,
  framexleftmargin=10pt,
  tabsize=2
}
\usepackage{color}
\usepackage{dcolumn} % decimal align in tables
\usepackage{bm} % bold math
\usepackage{graphicx}
\usepackage{multirow} % for table cells to span rows
\usepackage{pifont} % for checkmarks
\usepackage{amsmath} % matrix
\usepackage{float}
\usepackage{booktabs}
\usepackage{tabularx}
\usepackage{natbib}
\usepackage{gensymb}
\usepackage{rotating}
\usepackage{threeparttable}
\usepackage{comment}
\usepackage[normalem]{ulem}

\usepackage{xspace}
\newcommand{\pq}{\texttt{p$^\dagger$q}\xspace}
\newcommand{\np}{\textsc{NumPy}}

\begin{document}
  
\author{A. Eugene DePrince III}
\email{adeprince@fsu.edu}
\affiliation{
             Department of Chemistry and Biochemistry,
             Florida State University,
             Tallahassee, FL 32306-4390, USA}
\affiliation{
             FSU Quantum Initiative,
             Florida State University,
             Tallahassee, FL 32306-4390, USA}
\affiliation{
             Google Quantum AI, 
             San Francisco, CA, USA} 
             
\author{Nicholas C. Rubin}
\affiliation{
             Google Quantum AI, 
             San Francisco, CA, USA}   

\title{Automated generation and execution of large-scale semidefinite programs for variational two-electron reduced density matrix theory}

\begin{abstract}
The variational two-electron reduced density matrix (v2RDM) method provides a wave-function-free approach to the description of strong electron correlation whose computational cost scales polynomially with the number of correlated orbitals. Here, we present an automated pipeline for the generation and execution of semidefinite programs (SDPs) associated with the v2RDM problem. Algebraic constraints encoding ensemble $N$-representability conditions are generated via the \pq second-quantized algebra package, and the \pq-generated expressions are automatically translated into a sparse-matrix representation that can be digested by an appropriate SDP solver. We have implemented a boundary-point SDP (BPSDP) solver in the \texttt{libsdp} package that can execute the v2RDM SDPs on either CPU or GPU hardware. We apply the code in several contexts. First, two-particle conditions are used to describe the 76-orbital model of the iron–molybdenum cofactor (FeMoco) of nitrogenase, the simulation of which requires roughly 25 hours of compute time on a single Nvidia H100 GPU. Second, the performance of high-order (four-particle positivity) conditions is evaluated on a challenging two-dimensional H$_{10}$ cluster problem. Lastly, new classes of positivity constraints for mixed fermion-boson systems are generated and applied within the context of {\em ab initio} cavity quantum electrodynamics simulations. 

\end{abstract}

\maketitle

\section{Introduction}

The electronic energy for a system of electrons interacting by at most pairwise interactions is a linear functional of the two-electron reduced density matrix (2RDM). An energy minimization performed over the space of 2RDMs is, in principle, exact, provided that one can restrict this space to include only those 2RDMs that are derivable from an ensemble of physical $N$-electron density matrices, {\em i.e.}, those 2RDMs that are $N$-representable.\cite{Coleman63_668} 
Enforcing exact $N$-representability is not feasible in general, though, so any practical variational 2RDM (v2RDM) optimization must be carried out under a necessary, yet insufficient, subset of $N$-representability conditions. As the set of approximately $N$-representable 2RDMs contains the $N$-representable set, the v2RDM energy is underconstrained and yields a lower bound to the exact energy computed within the same single-particle basis set ({\em i.e.}, the full configuration interaction [CI] energy). The lower-bound nature of the approach is also reflected in the dual of the v2RDM problem, which seeks the largest constant $E$ such that the shifted Hamiltonian, $H - E$, is expressible as a Hermitian sum of squares (SOS) of the generating algebra.\cite{DePrince26_8751} 

Enforcing approximate $N$-representability conditions in v2RDM calculations can be useful in some settings. For example, v2RDM-based calculations that enforce two-positivity constraints (called PQG or DQG)\cite{Percus64_1756,Fujisawa01_8282} can be carried out with active spaces of more than 50 electrons in more than 50 orbitals,\cite{Mazziotti08_134108, Mazziotti10_014104, DePrince16_423, DePrince16_2260, DePrince19_276, DePrince19_6164} which is beyond the limits of exact CI approaches,\cite{deJong17_184111,Yoshida24_1185} and they provide qualitatively correct descriptions of static correlation in systems such as polyacene molecules.\cite{Mazziotti08_134108,Mazziotti11_5632,DePrince19_6164} However, a lack of strict $N$-representability leads to a variety of problems, including inconsistent descriptions of maximal and non-maximal spin- and angular-momentum projection states\cite{Ayers12_014110,DePrince19_032509,DePrince22_5966} and incorrect dissociation limits for heteronuclear diatomic systems.\cite{Ayers09_5558, Bultinck10_114113, Cooper11_054115} The latter issue traces to the absence of a derivative discontinuity in the energy for fractionally charged systems, analogous to the delocalization error associated with approximate density functionals.\cite{Yang08_792} Beyond these fundamental limitations, v2RDM can also display energy errors that call into question its usefulness for general chemical applications. As an example, Stair and Evangelista\cite{Evangelista20_104108} found that v2RDM under the PQG conditions displays energy errors that approach 200 mE$_\text{h}$ for two- and three-dimensional H$_{10}$ clusters. They also found that large errors (approaching 50 mE$_{\rm h}$) persist in some cases, even when the 2RDM satisfies partial three-particle conditions (specifically, the T2 condition\cite{Erdahl78_697,Percus04_2095}). Even the full set of three-positivity conditions (3POS) can result in energy errors on the order of 20--30 mE$_\text{h}$ in these systems,\cite{DePrince21_174110} which are disappointingly large, given the steep $\mathcal{O}(n^9)$ computational cost  associated with enforcing the 3POS conditions (where $n$ is the number of correlated spatial molecular orbitals).

These results raise an obvious question: which additional $N$-representability conditions are needed to remedy these deficiencies, and can such conditions be identified and imposed at a cost below that of the full 3POS conditions? Along these lines, several partial three-particle conditions have been proposed, including the T1, T2, and T2$'$ conditions.\cite{Erdahl78_697,Percus04_2095,Mazziotti06_032501} Subsets of four-particle conditions have also been explored, including those on diagonal elements,\cite{Davidson07_443} in the $(2,q)$ hierarchy,\cite{Mazziotti12_062507} and within the seniority-zero sector.\cite{Lain19_154104} However, a comprehensive evaluation regarding which conditions matter for specific classes of problems, as well as the convergence rate of the $p$-positivity hierarchy in real chemical applications, has yet to be performed. A practical obstacle to such a study is the implementation barrier. The linear maps that relate the various hole- and particle-hole-type RDMs (or combinations thereof) to the particle-type RDMs must be derived from the anticommutation relations for fermionic creation and annihilation operators, resolved into spin sectors, and implemented efficiently, and the resulting expressions become increasingly complex with the rank of the matrices. As an example, there are dozens of distinct linear maps across the three- and four-positivity conditions. Deriving and implementing these maps is tedious and error-prone, and it discourages the exploratory work needed to answer the questions posed above.

Many-body electronic structure theorists are familiar with this type of implementation bottleneck.  The equations of high-order coupled-cluster (CC) theory,\cite{Coester58_421,Kuemmel60_477,Cizek66_4256,Cizek69_35,Shavitt72_50,Li99_1,Musial07_291}  equation-of-motion (EOM) CC\cite{Emrich81_379,Bartlett89_57,Bartlett93_7029} theory, and related many-body methods are also complex, and their derivation and implementation have been largely automated by computer algebra tools that apply Wick's theorem or the anticommutation relations of second-quantized operators directly. Examples include the Tensor Contraction Engine,\cite{Hirata03_9887} SecondQuantizationAlgebra (SQA),\cite{Chan09_124102} SeQuant,\cite{Valeev26_142502} Wick\&d,\cite{Evangelista22_064111} and \pq.\cite{DePrince21_e1954709,DePrince25_6679} To our knowledge, no comparable framework exists for generating equations and code for $N$-representability conditions in v2RDM theory. In this work, we describe such a framework, within the \pq package,\cite{deprince_pdaggerq} which pairs a C++ engine for second-quantized algebra with a Python interface. Each matrix appearing in the v2RDM semidefinite program (SDP) is specified only by a string of creation and annihilation operators and the form of the expectation value that defines its elements (a product or an anticommutator). From this specification, \pq generates the linear maps to the particle RDMs, along with spin-symmetry constraints\cite{Mazziotti05_052505,Ayers12_014110} and contraction constraints connecting particle RDMs of different ranks. The spin-sector and block structure of each matrix follow automatically, and the constraints are assembled directly into a sparse representation of the SDP, which can then be solved with an appropriate SDP algorithm. Generating new conditions on the 2RDM thus requires only a few lines of Python code, as opposed to a full derivation and implementation, making the systematic exploration of $N$-representability conditions more practical.

This work is outlined as follows. Section \ref{SEC:PRIMAL_DUAL_FORMULATION} sets up the primal and dual formulations of the v2RDM SDP. Section \ref{SEC:AUTO} then describes the automated symbolic generation of ensemble $N$-representability conditions, along with the numerical representation of the problem we use in our SDP optimization algorithm. The capabilities of the automated v2RDM pipeline are then showcased in Sec.~\ref{SEC:RESULTS}, where we consider the application of the PQG conditions to the 76-orbital model of the iron–molybdenum cofactor (FeMoco) of nitrogenase, the role of high-order conditions (four-particle positivity, 4POS) for the challenging two-dimensional H$_{10}$ cluster problem, and the generation of new classes of positivity constraints for mixed fermion-boson systems ({\em e.g.}, for {\em ab initio} cavity quantum electrodynamics [QED]\cite{DePrince23_041301} applications). Some concluding remarks are presented in Sec.~\ref{SEC:CONCLUSIONS}.

\section{Primal-Dual v2RDM SDP Formulation}
\label{SEC:PRIMAL_DUAL_FORMULATION}

The electronic energy for a non-relativistic many-electron system is
\begin{align}
\label{EQN:ENERGY}
    E = \sum_{pq} h_{p q} {}^1D^{p}_{q} + \frac{1}{2} \sum_{pqrs} (p r|q s) {}^2D^{p q}_{r s}
\end{align}
where $h_{p q}$ and $(p r |q s )$ represent elements of the core Hamiltonian matrix and a two-electron repulsion integral in chemists' notation, respectively. The symbols ${}^1D^{p}_{q}$ and ${}^2D^{p q}_{r s}$ are elements of the 1RDM and 2RDM, respectively, and are defined as
\begin{align}
    \label{EQN:1RDM}
    {}^1D^{p}_{q} &= \langle  a^\dagger_{p} a_{q} \rangle \\
    \label{EQN:2RDM}
    {}^2D^{p q}_{r s} &= \langle a^\dagger_{p} a^\dagger_{q} a_{s} a_{r} \rangle
\end{align}
Here, $a^\dagger_{p}$ and $a_{p}$ are fermionic operators that create and annihilate electrons in spin orbital $\phi_{p}$. The goal of the v2RDM approach is to minimize Eq.~\ref{EQN:ENERGY} with respect to the elements of the 1RDM and 2RDM, while imposing necessary conditions for ensemble-state $N$-representability; an ensemble-state $N$-representable RDM is one that is derivable from an ensemble of physical $N$-electron density matrices. These $N$-representability conditions generally take four forms: (i) antisymmetry constraints that reflect the fermionic anticommutation relations (${}^2D^{p q}_{r s} = -{}^2D^{q p}_{r s} = -{}^2D^{p q}_{s r} = {}^2D^{q p}_{s r}$), (ii) expectation value constraints that encode fundamental symmetries of the system ({\em e.g.}, particle number,  spin projection, etc.), (iii) contraction constraints that relate the elements of the 2RDM to the 1RDM (or the elements of the 3RDM to the 2RDM, etc.), and (iv) constraints on the eigenvalues of the 2RDM (${}^2{\bf D}\succeq 0$), as well as other RDMs that are related to the 2RDM via contraction or linear maps implied by the fermionic anticommutation relations. This last class of constraints makes the v2RDM optimization a semidefinite programming (SDP) problem.

Let us collect our RDMs into a single quantity ${\bf D} = \{{}^1{\bf D}, {}^2{\bf D}, ...\}$, where the ellipsis represents additional one-, two-, three-body, etc. RDMs that will be introduced below. Now, the primal form of the v2RDM SDP can be expressed as
\begin{align}
\label{EQN:PRIMAL}
\begin{split}
\min_{\mathbf{D}} \quad & \langle \mathbf{H},\mathbf{D}\rangle \\
\text{such that} \quad & \mathcal{A}(\mathbf{D}) = \mathbf{b} \\
\text{and} \quad & M(\mathbf{D}) \succeq 0,
\end{split}
\end{align}
where ${\bf H}$ is the Hamiltonian integral tensor containing the 1- and 2-electron integrals, arranged such that the inner product $\langle {\bf H},{\bf D}\rangle$ recovers Eq.~\ref{EQN:ENERGY}, and $\mathcal{A}$ is a linear mapping from the space of RDMs to the constraint space $\mathbb{R}^m$, and $\mathbf{b}$ contains the constraint target values ({\em e.g.}, electron number $\eta$, spin projections, etc.). The linear mapping $M(\mathbf{D})$ constructs a block-diagonal matrix from tensors in ${\bf D}$
\begin{equation}
\label{EQN:MAPPING}
M({\bf D})=\left(\begin{array}{cccc}
                           {}^1{\bf D}&0&0\\
                           0&{}^2{\bf D}&0\\
                           0&0&...\\
                           \end{array}\right) 
\end{equation}
In practice, each matrix in $M(\mathbf{D})$ is itself block diagonal, owing to spin symmetry (see Sec.~\ref{SEC:SPIN}), and the blocks of $M(\mathbf{D})$ are actually these smaller-sized spin blocks. Positive semidefiniteness is imposed on each block separately.
The dual SDP is a complementary problem to Eq.~\ref{EQN:PRIMAL} and is given by
\begin{align}
\label{EQN:DUAL}
\begin{split}
\max_{\mathbf{y},\,\mathbf{Z}} \quad & \mathbf{b}^T \mathbf{y} \\
\text{such that} \quad & \mathcal{A}^*(\mathbf{y}) + M^*(\mathbf{Z}) = \mathbf{H} \\
\text{and} \quad & \mathbf{Z} \succeq 0,
\end{split}
\end{align}
Here, $\mathbf{y} \in \mathbb{R}^m$ is a vector of dual variables, which are the Lagrange multipliers that correspond to the equality constraints $\mathcal{A}(\mathbf{D}) = \mathbf{b}$, and $\mathbf{Z} = \mathrm{diag}({}^1\mathbf{Z}, {}^2\mathbf{Z}, \dots) \succeq 0$ is the block-diagonal dual matrix of multipliers enforcing $M(\mathbf{D}) \succeq 0$. The mappings $\mathcal{A}^*$ and $M^*$ are the adjoints of $\mathcal{A}$ and $M$, respectively, which map the dual multipliers $\mathbf{y}$ and dual matrices $\mathbf{Z}$ back into the RDM tensor space.

We use the boundary-point SDP (BPSDP) algorithm,\cite{Wiegele06_277,Wiegele09_336,Mazziotti11_083001} to solve the primal-dual SDP pair (Eqs.~\ref{EQN:PRIMAL} and \ref{EQN:DUAL}). We begin with the augmented Lagrangian function for the dual problem
\begin{align}
\label{EQN:AUG_LAGRANGIAN}
\mathcal{L}(\mathbf{y}, \mathbf{Z}; \mathbf{D}) &= \mathbf{b}^T \mathbf{y} + \langle \mathbf{D}, \mathbf{H} - \mathcal{A}^*(\mathbf{y}) - M^*(\mathbf{Z}) \rangle \nonumber \\
&- \frac{1}{2\mu} \left\| \mathbf{H} - \mathcal{A}^*(\mathbf{y}) - M^*(\mathbf{Z}) \right\|_F^2,
\end{align}
which should be maximized with respect to ${\bf y}$ and ${\bf Z}$. Here,  $\mu > 0$ is a penalty parameter, and $\|\cdot\|_F$ denotes the Frobenius norm over the RDM tensor space. At step $k+1$, an updated dual vector, ${\bf y}^{k+1}$ can be obtained from the linear system that derives from the stationary condition $\nabla_{\mathbf{y}}\mathcal{L} = 0$:
\begin{equation}
\label{EQN:CG}
\mathcal{A}\mathcal{A}^*(\mathbf{y}^{k+1}) = \mathcal{A}\left(\mathbf{H} - M^*(\mathbf{Z}^k)\right) + \mu\left(\mathbf{b} - \mathcal{A}(\mathbf{D}^k)\right)
\end{equation}
This equation can be solved for $\mathbf{y}^{k+1}$ via the conjugate gradient (CG) algorithm, the computational cost of which is dominated by repeated evaluations of the left-hand side operator $\mathcal{A}\mathcal{A}^*$ [the action of the adjoint map $\mathbf{u} = \mathcal{A}^*(\mathbf{y})$, followed by the forward map $\mathcal{A}(\mathbf{u})$]. Given the updated dual vector $\mathbf{y}^{k+1}$, revised estimates for $\mathbf{D}^{k+1}$ and $\mathbf{Z}^{k+1}$ are obtained from the block-wise eigendecomposition of $\mathbf{U} = M\left(\mu \mathbf{D}^k + \mathcal{A}^*(\mathbf{y}^{k+1}) - \mathbf{H}\right)$:
\begin{align}
    \label{EQN:UPDATE_XZ}
    \mathbf{D}^{k+1} &= \frac{1}{\mu} M^*(\mathbf{U}_+) \\
    \mathbf{Z}^{k+1} &= -\mathbf{U}_-
\end{align}
where $\mathbf{U}_+$ and $\mathbf{U}_-$ denote the positive and negative spectral projections of $\mathbf{U}$, respectively. Every $n_\mu$ iterations, the penalty parameter is rescaled by the ratio of the primal and dual errors, $\mu \leftarrow \mu\, \|\mathcal{A}(\mathbf{D}) - \mathbf{b}\| / \|\mathcal{A}^*(\mathbf{y}) + M^*(\mathbf{Z}) - \mathbf{H}\|$, which places more weight on either the primal or dual feasibility, depending on which is larger. 

\section{Automated Construction of the v2RDM SDP}
\label{SEC:AUTO}

\subsection{Positivity conditions from operator generators}
\label{SEC:GENERATORS}

Consider an operator $C = \sum_I c_I O_I$, where each $O_I$ is a product of a fixed number of creation and/or annihilation operators with a specific form, {\em i.e.}, having the same sequence of creators and annihilators, while differing only in their orbital labels. Necessary conditions for ensemble $N$-representability can be derived from the fact that the expectation value of the Hermitian square of this operator is non-negative:
\begin{equation}
\label{EQN:NORM_SQUARED}
\langle CC^\dagger \rangle = \sum_{I,J} c_I c_J^* \langle O_I O_J^\dagger \rangle \ge 0.
\end{equation}
Equation~\ref{EQN:NORM_SQUARED} must hold for any choice of complex coefficients $\mathbf{c}$, which implies that the matrix with elements $B^I_J = \langle O_I O_J^\dagger\rangle$ is positive semidefinite (PSD).  As an example, the choice $O_I = a^\dagger_i a^\dagger_j$ and $O_J = a^\dagger_a a^\dagger_b$ (so $O_J^\dagger = a_b a_a$) recovers the definition for the elements of the 2RDM in Eq.~\ref{EQN:2RDM} and implies that the 2RDM must be PSD, while other choices imply the non-negativity of other RDMs. In the context of the SDP for the v2RDM problem, the positivity constraints generated with different choices for $O_I$ and $O_J^\dagger$ place restrictions on the 2RDM in the form of linear mappings between the 2RDM and other PSD RDMs; the precise form of the mappings is dictated by the anticommutation relations of the operators comprising each $O_I$ and $O_J^\dagger$, as well as contraction relations connecting the $(k+1)$- and $k$-electron RDMs. Table \ref{TAB:POSITIVITY} gives the definitions for the elements of up to the four-body RDMs and introduces the $x/o$ notation of Ref.~\citenum{Mazziotti12_263002}, which we use to label different RDMs comprising the primal solution, ${\bf D}$. The symbols $x$ and $o$ refer to creation and annihilation operators, respectively, appearing in each $O_I$. Additional positivity conditions can be generated from anticommutators of $O_I$ and $O_J^\dagger$, {\em i.e.}, from the non-negativity of matrices defined by
\begin{align}
    B^I_J = \langle O_I O_J^\dagger + O_J^\dagger O_I\rangle
\end{align}
In this class, we have the T1 and T2 conditions,\cite{Erdahl78_697, Percus04_2095} defined by  anticommutators of the form $\{xxx,(xxx)^\dagger\}$ and $\{xxo,(xxo)^\dagger\}$, respectively. Unlike the Hermitian square $\langle O_I O_J^\dagger\rangle$, which depends on the 3RDM when $O_I$ is cubic, the anticommutator forms of T1 and T2 cause the three-body contributions to vanish, so these matrices are expressible only in terms of the 1RDM and 2RDM. In addition, we have the
T2$'$ condition, which  generalizes T2 by adding a one-body component to the operator. If $C = \sum_{ijk} x_{ijk}\, a^\dagger_i a^\dagger_j a_k$ and $C' = C + \sum_l y_l\, a^\dagger_l$, then the condition $\langle C'C'^\dagger\rangle + \langle C^\dagger C\rangle \ge 0$ for all complex $\mathbf{x}$ and $\mathbf{y}$ requires the matrix
\begin{equation}
\mathbf{B} =
\left(\begin{array}{cc}
\langle \{xxo,\,(xxo)^\dagger\} \rangle & \langle xxo\,(x)^\dagger \rangle \\[4pt]
\langle x\,(xxo)^\dagger \rangle & \langle x\,(x)^\dagger \rangle
\end{array}\right)
\end{equation}
to be PSD. As such, T2$'$ mixes an anticommutator (the three-body part) with ordinary Hermitian-square elements (the off-diagonal and one-body parts). 
\begin{table}[!htpb]
    \caption{Role strings and elements of the RDMs involving up to four-body operators.}
    \label{TAB:POSITIVITY}
    \begin{center}
    \setlength\extrarowheight{3pt}
    \begin{tabular}{l l l}
    \hline\hline
    role string & RDM name & RDM elements \\
    \hline
        $1$    & zero-body (normalization) & ${}^0D = \langle 1 \rangle = 1$ \\
    $x$    & one-electron            & ${}^1D^{i}_{a} = \langle a^\dagger_{i} a_{a} \rangle$ \\
    $o$    & one-hole                & ${}^1Q^{i}_{a} = \langle a_{i} a^\dagger_{a} \rangle$ \\
    $xx$   & two-electron            & ${}^2D^{ij}_{ab} = \langle a^\dagger_{i} a^\dagger_{j} a_{b} a_{a} \rangle$ \\
    $xo$   & electron-hole           & ${}^2G^{ij}_{ab} = \langle a^\dagger_{i} a_{j} a^\dagger_{b} a_{a} \rangle$ \\
    $oo$   & two-hole                & ${}^2Q^{ij}_{ab} = \langle a_{i} a_{j} a^\dagger_{b} a^\dagger_{a} \rangle$ \\
    $xxx$  & three-electron          & ${}^3D^{ijk}_{abc} = \langle a^\dagger_{i} a^\dagger_{j} a^\dagger_{k} a_{c} a_{b} a_{a} \rangle$ \\
    $xxo$  & two-electron-one-hole   & ${}^3E^{ijk}_{abc} = \langle a^\dagger_{i} a^\dagger_{j} a_{k} a^\dagger_{c} a_{b} a_{a} \rangle$ \\
    $xoo$  & one-electron-two-hole   & ${}^3F^{ijk}_{abc} = \langle a^\dagger_{i} a_{j} a_{k} a^\dagger_{c} a^\dagger_{b} a_{a} \rangle$ \\
    $ooo$  & three-hole              & ${}^3Q^{ijk}_{abc} = \langle a_{i} a_{j} a_{k} a^\dagger_{c} a^\dagger_{b} a^\dagger_{a} \rangle$ \\
    $xxxx$ & four-electron           & ${}^4D^{ijkl}_{abcd} = \langle a^\dagger_i a^\dagger_j a^\dagger_k a^\dagger_l a_d a_c a_b a_a \rangle$ \\
    $xxxo$ & three-electron-one-hole & ${}^4E^{ijkl}_{abcd} = \langle a^\dagger_i a^\dagger_j a^\dagger_k a_l a^\dagger_d a_c a_b a_a \rangle$ \\
    $xxoo$ & two-electron-two-hole   & ${}^4G^{ijkl}_{abcd} = \langle a^\dagger_i a^\dagger_j a_k a_l a^\dagger_d a^\dagger_c a_b a_a \rangle$ \\
    $xooo$ & one-electron-three-hole & ${}^4F^{ijkl}_{abcd} = \langle a^\dagger_i a_j a_k a_l a^\dagger_d a^\dagger_c a^\dagger_b a_a \rangle$ \\
    $oooo$ & four-hole               & ${}^4Q^{ijkl}_{abcd} = \langle a_i a_j a_k a_l a^\dagger_d a^\dagger_c a^\dagger_b a^\dagger_a \rangle$ \\
    \hline\hline
    \end{tabular}
    \end{center}
\end{table}

\subsection{Role strings, spin sectors, and PSD blocks}
\label{SEC:SPIN}
In our program, each matrix is specified by its role string(s) ({\em e.g.}, $xx$, $xo$, etc.) and its form, which is defined by the subset of role strings that appear within an anticommutator. Matrix elements between two anticommuted role strings are evaluated as $\langle \{O_I, O_J^\dagger\}\rangle$, and all other elements are products, $\langle O_I O_J^\dagger\rangle$. The matrices in Table~\ref{TAB:POSITIVITY} are all product-based, whereas the T1 and T2 matrices, for example, each involve a single anticommuted role string. All properties of the matrix follow from these identifiers. By convention, creators precede annihilators, so a role string consists of $n_x$ creators followed by $n_o$ annihilators. $O_I$ is the role string written from left to right, and $O_J^\dagger$ is the reversed string, with $x \leftrightarrow o$. For the two-electron-one-hole RDM (${}^3{\bf E}$), the role string is $xxo$, and we have $O_I = a^\dagger_i a^\dagger_j a_k$ and $O_J = a^\dagger_a a^\dagger_b a_c$ (so $O_J^\dagger = a^\dagger_c a_b a_a$). 
In this example, the creators-first convention fixes the choice of the $xxo$ representation of this RDM, as opposed to $xox$. Each $x$ or $o$ operator can carry either $\alpha$ or $\beta$ spin, and, within each group of operators with the same role, $\alpha$-spin operators precede $\beta$-spin operators. Each spin sector can be delineated by the number of $\alpha$-spin creators ($n_x^\alpha \in \{0,\dots,n_x\}$) and the number of $\alpha$-spin annihilators ($n_o^\alpha \in \{0,\dots,n_o\}$); a role string thus yields $(n_x+1)(n_o+1)$ distinct spin sectors. For the $xxo$ example, the role string generates six spin sectors: 
\begin{align}
    xxo \to  \{ &x_\alpha x_\alpha o_\alpha ,~x_\alpha x_\alpha o_\beta ,~x_\alpha x_\beta o_\alpha, \nonumber \\
    & x_\alpha x_\beta o_\beta ,~x_\beta x_\beta o_\alpha ,~ x_\beta x_\beta o_\beta \}
\end{align} 
The generators, $O_I$ and $O_J$, are antisymmetric with respect to exchange of the orbital labels of operators that share a role and a spin, so the row and column indices of the corresponding matrix only account for the ordered combinations $p<q<r<\cdots$ within each group ({\em e.g.}, $i<j$ [rows] and $a<b$ [columns] for $x_\alpha x_\alpha o_\alpha$). 
The generators have at most four sequences of operators that share both a role and a spin ($x_\alpha$, $x_\beta$, $o_\alpha$, $o_\beta$), so the dimension of a sector is
\begin{align}
d_s = \binom{n}{n_x^\alpha}\binom{n}{n_x-n_x^\alpha}\binom{n}{n_o^\alpha}\binom{n}{n_o-n_o^\alpha}
\end{align}
where $n$ is the number of spatial molecular orbitals. 

With spin labels, the role strings also dictate which sectors belong to the same PSD block. Each $O_I$ changes the numbers of $\alpha$- and $\beta$-spin electrons by fixed amounts
\begin{align}
    (\Delta N_\alpha, \Delta N_\beta) = (n_x^\alpha - n_o^\alpha,\; n_x^\beta - n_o^\beta)
\end{align}
where $n_x^\beta = n_x - n_x^\alpha$ and $n_o^\beta = n_o - n_o^\alpha$. For an ensemble of states with definite $N_\alpha$ and $N_\beta$ (as assumed throughout), the element $\langle O_I O_J^\dagger\rangle$ can be nonzero only if $O_I O_J^\dagger$ conserves $N_\alpha$ and $N_\beta$ separately, which requires $O_I$ and $O_J$ to have the same $(\Delta N_\alpha, \Delta N_\beta)$. For example, for the role string $xxo$, the $\alpha\alpha\alpha$ and $\alpha\beta\beta$ spin sectors couple because they both have $(\Delta N_\alpha, \Delta N_\beta) = (+1,0)$. Accounting for the coupling among other spin sectors yields the full spin-block structure of the ${}^3{\bf E}$ matrix:
\begin{equation}
{}^3{\bf E} =
\begin{pmatrix}
 {}^3{\bf E}^{\alpha\alpha\alpha}_{\alpha\alpha\alpha} & {}^3{\bf E}^{\alpha\alpha\alpha}_{\alpha\beta\beta} & 0 & 0 & 0 & 0 \\
 {}^3{\bf E}^{\alpha\beta\beta}_{\alpha\alpha\alpha} & {}^3{\bf E}^{\alpha\beta\beta}_{\alpha\beta\beta} & 0 & 0 & 0 & 0 \\
 0 & 0 & {}^3{\bf E}^{\alpha\beta\alpha}_{\alpha\beta\alpha} & {}^3{\bf E}^{\alpha\beta\alpha}_{\beta\beta\beta} & 0 & 0 \\
 0 & 0 & {}^3{\bf E}^{\beta\beta\beta}_{\alpha\beta\alpha} & {}^3{\bf E}^{\beta\beta\beta}_{\beta\beta\beta} & 0 & 0 \\
 0 & 0 & 0 & 0 & {}^3{\bf E}^{\alpha\alpha\beta}_{\alpha\alpha\beta} & 0 \\
 0 & 0 & 0 & 0 & 0 & {}^3{\bf E}^{\beta\beta\alpha}_{\beta\beta\alpha}
\end{pmatrix}
\end{equation}
The dimension of each PSD block is given by the cumulative dimension of sectors that couple within the block
\begin{align}
    d_b = \sum_{\text{sectors } s \in b} d_s
\end{align}

The T2$'$ matrix illustrates the case of mixed forms. It is specified by two role strings ($xxo$ and $x$), of which only $xxo$ is anticommuted, so the anticommutator is used between $xxo$ sectors, and the product is used otherwise. Spin-sector generation applies to each string separately, and the sectors from both strings are then grouped by $(\Delta N_\alpha, \Delta N_\beta)$ as in the other matrices. In general, if $C' = \sum_I c_I O_I$ runs over all of the role strings, and $C$ is the part of $C'$ built from the anticommuted role strings, the resulting matrix represents $\langle C'C'^\dagger\rangle + \langle C^\dagger C\rangle$, which is non-negative for any $N$-electron state. Any choice of role strings and anticommuted subset therefore yields a valid positivity condition.

Before moving on, we also note that Table \ref{TAB:POSITIVITY} defines the role string $1$, which represents the identity operator. On its own, this role string generates only the trivial normalization, ${}^0D = \langle 1 \rangle = 1$, but it is nonetheless potentially useful. For example, because it changes neither $N_\alpha$ nor $N_\beta$, the identity can be grouped with other number-conserving role strings (although we do not consider this possibility here). Moreover, as will be discussed in Sec.~\ref{SEC:QED_GENERATORS}, the photon-dressed form of the identity generates the photon RDM, which will be relevant in cavity QED applications.

\subsection{Symbolic constraint generation}
\label{SEC:SYMBOLIC}

The first type of constraints that are automatically generated by \pq are the linear maps that relate the spin sectors of the hole- and particle-hole-type RDMs in Table~\ref{TAB:POSITIVITY} (or the T1, T2, and T2$^\prime$ matrices) to the appropriate spin sectors of the $k$-body particle-type RDMs, ${}^k{\bf D}$.  As an example, let us consider the ${}^3{\bf E}^{\alpha\alpha\alpha}_{\alpha\beta\beta}$ spin sector pair defined by the spin-dependent role strings $x_\alpha x_\alpha o_\alpha$ (for $O_I$) and $x_\alpha x_\beta o_\beta$ (for $O_J$). The role strings and spin labels yield the following operators with explicit symbolic indices: $O_I = a^\dagger_{i_\alpha} a^\dagger_{j_\alpha} a_{k_\alpha}$ and $O^\dagger_J = a^\dagger_{c_\beta} a_{b_\beta} a_{a_\alpha}$. Here, we have introduced spin labels on the creation and annihilation operators, so $a^\dagger_{i_\alpha}$ should be understood as a creator acting on spatial orbital $i$ with spin label $\alpha$. In general, $O_I$ and $O_J$ are processed by \pq as a product or an anticommutator, depending on the matrix's form; for ${}^3{\bf E}$, specifically, the form is a product. \pq brings the string(s) of operators to normal order with respect to the true vacuum state and returns the matrix element, $B^I_J$, as a list of products of spin-labeled $k$-body RDM elements, Kronecker deltas, and scalars.
Each such list maps onto a linear constraint of the form
\begin{equation}
\label{EQN:LINEAR_CONSTRAINT}
\sum _k \sum_{K,L} c^{IJ}_{KL}\, {}^kD^K_L - B^I_J = -s^{IJ}
\end{equation}
Here, $s^{IJ}$ are scalar terms, which are collected in the constraint vector, $\mathbf{b}$ ({\em cf.} Eq.~\ref{EQN:PRIMAL}), and the Kronecker deltas are absorbed into the coefficients, $c^{IJ}_{KL}$. 

Additional contraction and expectation value constraints can be generated with the same machinery, with appropriate operator expressions. For example, the trace constraint on the 2RDM is
\begin{align}
\label{EQN:TRACE}
            0 &= - \eta_\sigma \eta_\tau + \langle {N}_\sigma N_\tau \rangle ~\forall ~\sigma \le \tau
\end{align}
where $\eta_\sigma$ is the number of $\sigma$-spin electrons, and ${N}_\sigma$ is the number operator for $\sigma$-spin electrons
\begin{align}
    {N}_\sigma = \sum_p a^\dagger_{p_\sigma} a_{p_\sigma}
\end{align}
Similarly, we have contraction conditions relating the spin sectors of the $k$-particle and ($k+1$)-particle RDMs 
\begin{align}
\label{EQN:CONTRACTION}
            0 = - \eta_\sigma \langle O_IO^\dagger_J \rangle + \langle {N}_\sigma O_IO^\dagger_J \rangle
\end{align}
Here, $O_I$ (and its complement $O_J^\dagger$) can refer to any of the spin sectors of the all-particle role strings. Our code also implements an expectation value constraint on the square of spin
\begin{align}
    0 = -S(S+1) + \langle S_z \rangle + \langle S_z S_z \rangle  + \langle S_- S_+ \rangle
\end{align}
where $S = \frac{1}{2}(\eta_\alpha - \eta_\beta)$ is the total spin quantum number, assuming the maximal spin projection, and 
\begin{align}
    S_- &= \sum_p a^\dagger_{p_\beta} a_{p_\alpha} \\
    S_+ &= \sum_p a^\dagger_{p_\alpha} a_{p_\beta} \\
    S_z &= \frac{1}{2} \sum_p \left ( a^\dagger_{p_\alpha} a_{p_\alpha} - a^\dagger_{p_\beta} a_{p_\beta} \right ) 
\end{align}
We have also implemented constraints on the maximal spin projection of the form
\begin{align}
    0 &= \langle O_I S_+\rangle 
\end{align}
where $O_I$ is restricted to the $x_\beta o_\alpha$ spin sector of the role string $xo$. This constraint derives from the fact that, for a state with maximal spin projection, $S_+|\Psi\rangle = 0$, so $\langle O_I S_+\rangle = 0$ for any $O_I$. We choose the $x_\beta o_\alpha$ sector  so that $O_I S_+$ conserves $N_\alpha$ and $N_\beta$.\cite{Ayers12_014110} Note that a user can choose not to impose spin constraints, in which case $N_\sigma$ and $\eta_\sigma$ in Eq.~\ref{EQN:CONTRACTION} are replaced by
the total number operator and total number of electrons, $\eta_\alpha + \eta_\beta$, and the $S^2$ and spin-projection constraints are omitted.
In all of these cases, as was done for the linear mapping constraints, the operator products are normal ordered and the expressions, which depend on the $k$-electron RDMs, are brought to the form of Eq.~\ref{EQN:LINEAR_CONSTRAINT}.  

\subsection{Numerical representation of the SDP}
\label{SEC:NUMERICAL}

We have implemented our SDP generator in Python. The \pq-generated expressions are evaluated with vectorized \np operations and assembled into sparse matrices with SciPy. The BPSDP solver in \texttt{libsdp}\cite{deprince_libsdp}  is implemented in JAX, which allows the same code to run on CPU or GPU hardware.
SDPs involving different $N$-representability conditions are generated through a single Python function. As an example, the following code snippet sets up SDPs involving the PQG and T2 or T2$'$ conditions:
\begin{lstlisting}
from pdaggerq.numerical.codegen.auto_v2rdm import v2rdm_sdp

sdp_t2  = v2rdm_sdp(nalpha, nbeta, nmo, 
            oei, tei, q2=True, g2=True, t2=True)
sdp_t2p = v2rdm_sdp(nalpha, nbeta, nmo, 
            oei, tei, q2=True, g2=True, t2_prime=True)
\end{lstlisting}
\noindent Here, \texttt{nalpha} and \texttt{nbeta} are the numbers of $\alpha$- and $\beta$-spin electrons, \texttt{nmo} is the number of spatial orbitals, and \texttt{oei} and \texttt{tei} are the one- and two-electron integrals. Equivalently, the T2 and T2$'$ matrices could be defined via custom blocks, which are specified by the complete set of role strings and the subset of role strings that enter an anticommutator (Sec.~\ref{SEC:SPIN}):
\begin{lstlisting}
sdp_t2  = v2rdm_sdp(nalpha, nbeta, nmo,
            oei, tei, q2=True, g2=True, 
            blocks={'MYT2': ('xxo', ('xxo',))})
sdp_t2p = v2rdm_sdp(nalpha, nbeta, nmo, 
            oei, tei, q2=True, g2=True,
            blocks={'MYT2P': (('xxo', 'x'), ('xxo',))})
\end{lstlisting}
\noindent Both forms generate identical SDPs. New conditions of this type, such as a T1$'$ condition that adds a one-body component to T1, can be defined in the same way ({\em i.e.}, \texttt{blocks=\{'T1P': (('xxx', 'x'), ('xxx',))\}}), without modifying the generator. Constraints that depend on external information can also be appended to the SDP after it is generated, either as additional rows of the constraint matrix (for equality constraints) or as additional PSD blocks linked to the RDMs by such rows (for inequality constraints). For example, an independently determined bound $\kappa$ on the eigenvalues of the 1RDM, $\kappa\mathbf{I} - {}^1\mathbf{D} \succeq 0$, can be imposed by adding a PSD block $\mathbf{S}$ and the linear constraints $\mathbf{S} + {}^1\mathbf{D} = \kappa \mathbf{I}$.

In the remainder of this section, we describe the layout of the primal and dual variables, as well as the constraint matrix produced by the generator. The primal solution is stored as a single vector, $\mathbf{x}$, formed by concatenating the flattened blocks of $M(\mathbf{D})$, which, themselves, are dense $d_b \times d_b$ matrices in row-major order; $\mathbf{x}$ has total length $\sum_b d_b^2$. Within block $b$, an element in row sector $s$ and column sector $s'$, with composite indices $I$ and $J$, respectively, sits at the $\left [ o_b + (o_s + I)\,d_b + (o_{s'} + J)\right ]$-th position in $\mathbf{x}$. Here, $o_b$ is the offset of the block within $\mathbf{x}$, and $o_s$ is the offset of sector $s$ within the block. 

We represent the linear map, $\mathcal{A}$, as a sparse matrix, $\mathbf{A}$, in compressed sparse row (CSR) format. Each row of ${\bf A}$ corresponds to a constraint, and each column corresponds to an element of $\mathbf{x}$. This representation allows us to write
\begin{align}
    \mathcal{A}(\mathbf{D}) &= \mathbf{A}\mathbf{x} \\
    \mathcal{A}^*(\mathbf{y}) &= \mathbf{A}^T\mathbf{y}
\end{align}
The Hamiltonian is stored as a vector, $\mathbf{c}$, such that
\begin{align}
    \langle \mathbf{H},\mathbf{D}\rangle = \mathbf{c}^T\mathbf{x}
\end{align}
The energy expression is generated via the same procedure as the expressions for constraints (the rows of $\mathbf{A}$), {\em i.e.}, via \pq's algebraic engine, with built-in operators that represent the one- and two-electron parts of the Hamiltonian.

Each symbolic expression of the form of Eq.~\ref{EQN:LINEAR_CONSTRAINT} ({\em e.g.}, one corresponding to a pair of spin sectors in a given block) defines a family of constraints. Within a family, each row and column label pair ($I \le J$ for diagonal sector pairs; see below) represents a single constraint. Rather than looping over these individual constraints, we evaluate each term in the expression for many constraints at once. We generate the values of the free indices (the row and column labels) in chunks, and, for each term, we construct an integer grid that spans the constraints in the chunk and all values of the term's summed indices. The coefficient of each term, multiplied by any Kronecker deltas the term contains, is evaluated on this grid, and we discard entries of the grid that vanish due to the deltas. As mentioned above, scalar terms (which do not involve RDM elements) are folded into $\mathbf{b}$. The orbital labels of each non-zero entry are then mapped to a position in $\mathbf{x}$. This mapping requires careful logic regarding signs. Although \pq's algebraic engine returns RDM elements with canonically-ordered spin labels, the numerical values of the orbital labels within a group of operators sharing a role and a spin are not necessarily ordered at a given grid point. For example, a summed index is not restricted relative to the free indices in the same group, and free indices that originate in different groups of $B^I_J$ can appear in the same group of an RDM element. As such, we must adjust the sign of each entry to reflect the parity of the permutation that sorts the labels into $i<j<\cdots$ (or $a<b<\cdots$) order. Moreover, entries in which two such labels coincide vanish by antisymmetry and are thus discarded. We assemble the resulting (row, column, value) triplets into a CSR matrix for each chunk, summing repeated contributions to the same element, and the chunks are stacked vertically to form $\mathbf{A}$. The chunk size bounds the working memory, and no Python loops over individual matrix elements are required.

Finally, assuming real orbitals, every block is real symmetric, so the constraints associated with $B^I_J$ and $B^J_I$ are redundant. Our program therefore only generates constraints for the upper triangle ($I \le J$) of each diagonal sector pair and for a single ordered pair of distinct sectors within each block. 
Each coefficient is split evenly between the $(I,J)$ and $(J,I)$ elements as the triplets are generated.
Alternatively, because the primal blocks are symmetric, each off-diagonal pair can be stored only once, as the upper-triangle element with the full coefficient. This ``folded'' representation, ${\bf A}_{\rm f}$, halves the number of non-zero elements in ${\bf A}$ and is exact for symmetric $\mathbf{x}$. The forward map is unchanged,
\begin{align}
\label{EQN:Af}
    \mathcal{A}(\mathbf{D}) = \mathbf{A}_{\rm f}\mathbf{x},
\end{align}
while the adjoint map is obtained by symmetrizing each block of $\mathbf{A}_{\rm f}^T\mathbf{y}$. The operator on the left-hand side of Eq.~\ref{EQN:CG} then becomes
\begin{align}
\label{EQN:AfWAf}
    \mathcal{A}\mathcal{A}^* = \mathbf{A}_{\rm f}\,\mathbf{W}\,\mathbf{A}_{\rm f}^T,
\end{align}
where $\mathbf{W}$ is a diagonal matrix whose elements are 1 for the diagonal elements of each block and 1/2 otherwise.
As a post-processing step, ${\bf A}$ is cleaned to remove any constraints that are trivially satisfied ({\em i.e.}, $0=0$).

\subsection{Extension to the cavity QED case}
\label{SEC:QED}

The machinery described above is easily extensible to the treatment of mixed fermion-boson systems. Here, we consider strong coupling between electronic degrees of freedom and a single quantized radiation mode associated with an optical cavity, within the framework of cavity quantum electrodynamics (QED). We treat the photon mode in a truncated photon-number basis, as has been done in previous work generalizing v2RDM theory for  cavity QED applications.\cite{DePrince22_053710}

\subsubsection{Hamiltonian and photon-number basis}
\label{SEC:QED_HAMILTONIAN}

Interactions between molecular degrees of freedom and a single optical cavity mode are captured by the Pauli-Fierz Hamiltonian,\cite{Spohn04_book,Rubio18_1} which we express in the length gauge and under the dipole approximation
\begin{align}
\label{EQN:PF_HAMILTONIAN}
H &= H_{\rm e} + \omega\, b^\dagger b
- \sqrt{\frac{\omega}{2}}\, \bm{\lambda}\cdot{\bm{\mu}}\left(b^\dagger + b\right) + \frac{1}{2}\left(\bm{\lambda}\cdot {\bm{\mu}}\right)^2 
\end{align}
The first and second terms are the isolated electronic and cavity mode Hamiltonians, respectively, where, in the second term, $\omega$ is the frequency of the cavity mode, and $b^\dagger$ and $b$ are bosonic creation and annihilation operators that act on the cavity mode. The third term is the bilinear coupling term, in which ${\bm{\mu}} = {\bm \mu}_\text{e} + {\bm \mu}_\text{nu}$ is the total molecular dipole operator (electronic plus nuclear), and $\bm{\lambda}$ is a vector that parametrizes the coupling between the light and matter degrees of freedom. The last term is the dipole self-energy (DSE) term, which contains no photon operators. The one- and two-electron parts of the DSE term can be folded into $H_\text{e}$, after which the electronic part of the energy can be expressed in terms of the 1RDM and 2RDM, as in Eq.~\ref{EQN:ENERGY}, but with appropriately dressed one- and two-electron integrals, as well as the constant term from the nuclear-only part of the DSE. Additional details regarding the implementation of the DSE term can be found in Ref.~\citenum{DePrince23_041301}.

To treat the photon-dependent parts of the energy, we begin by representing the photon mode in a truncated photon-number basis $\{|A^\text{p}\rangle\}$, with $A = 0, \dots, n_{\rm ph}$, where $n_{\rm ph}$ is the maximum photon number. Within this space, we can write $b^\dagger = \sum_{A=0}^{n_{\rm ph}-1} \sqrt{A+1}\, |(A+1)^\text{p}\rangle\langle A^\text{p}|$, and all photonic quantities can be expressed in terms of the photon density operators $|A^\text{p}\rangle\langle B^\text{p}|$, which commute with the fermionic creation and annihilation operators. The relevant photon- and electron-photon RDMs are then
\begin{align}
\label{EQN:PHOTON_RDM}
{}^0D^{A}_{B} &= \langle |A^\text{p}\rangle\langle B^\text{p}| \rangle \\
\label{EQN:EP_1RDM}
{}^1D^{pA}_{qB} &= \langle a^\dagger_p a_q\, |A^\text{p}\rangle\langle B^\text{p}| \rangle \\
\label{EQN:EP_2RDM}
{}^2D^{pqA}_{rsB} &= \langle a^\dagger_p a^\dagger_q a_s a_r\, |A^\text{p}\rangle\langle B^\text{p}| \rangle \\
\text{etc.} \nonumber
\end{align}
Because $\sum_A |A^\text{p}\rangle\langle A^\text{p}|$ is the identity within the truncated photon space, the photon RDM is normalized,
\begin{align}
\label{EQN:PHOTON_TRACE}
\sum_{A=0}^{n_\text{ph}} {}^0D^{A}_{A} = 1
\end{align}
which we impose as a constraint in the SDP.

With these electron-photon RDM definitions, the photon parts of the energy can then be constructed from the relationships
\begin{align}
\label{EQN:PHOTON_HAM_CONTRIBUTION}
    \langle b^\dagger b \rangle = \sum_{A=0}^{n_\text{ph}} A\, {}^0D^A_A
\end{align}
and
\begin{align}
\label{EQN:BLC_CONTRIBUTION}
     {\bm \lambda}\cdot\langle {\bm{\mu}} & \left (b^\dagger + b\right)\rangle = \nonumber \\ 
     &\sum_{A=0}^{n_{\rm ph}-1} \sqrt{A+1}\sum_{pq} d_{pq} \left ( {}^1D^{p,A+1}_{q,A} + {}^1D^{p,A}_{q,A+1}\right) \nonumber\\
    &+d_\text{nu}\sum_{A=0}^{n_{\rm ph}-1}\sqrt{A+1} \left ({}^0D^{A+1}_A + {}^0D^A_{A+1}\right)
\end{align}
where $d_{pq}$ is an element of the matrix representation of the ${\bm \lambda}$-weighted electronic dipole operator, and $d_\text{nu} = {\bm \lambda}\cdot{\bm{\mu}}_\text{nu}$ is the ${\bm \lambda}$-weighted nuclear part of the dipole operator (which is a scalar within the Born-Oppenheimer approximation). The bilinear coupling part of the energy is then $-\sqrt{\omega/2}$ times Eq.~\ref{EQN:BLC_CONTRIBUTION}.
Note that the purely electronic RDMs are related to the electron-photon RDMs by partial traces in the photon space, {\em e.g.}, ${}^2D^{pq}_{rs} = \sum_A {}^2D^{pqA}_{rsA}$.

\subsubsection{Photon-dressed generators}
\label{SEC:QED_GENERATORS}

We can generalize the positivity conditions of Sec.~\ref{SEC:GENERATORS} for the cavity QED case by augmenting each generator with a photon density operator. Consider the operator 
\begin{align}
    C = \sum_{I,A} c_{IA}\, O_I\, |A^\text{p}\rangle\langle 0^\text{p}|
\end{align}
We can write  
\begin{align}
|A^\text{p}\rangle\langle 0^\text{p}|\,\left(|B^\text{p}\rangle\langle 0^\text{p}|\right)^\dagger = |A^\text{p}\rangle\langle B^\text{p}|
\end{align}
in which case the cavity QED analog of Eq.~\ref{EQN:NORM_SQUARED} implies that the matrix with elements
\begin{align}
\label{EQN:DRESSED_MATRIX}
B^{IA}_{JB} = \langle O_I O_J^\dagger\, |A^\text{p}\rangle\langle B^\text{p}| \rangle
\end{align}
is PSD. Every Hermitian-square condition in Table~\ref{TAB:POSITIVITY} therefore has a photon-dressed counterpart, and the photon-dressed PQG, 3POS, and 4POS conditions follow with no additional derivation. 
The RDMs used in Ref.~\citenum{DePrince22_053710}, which are labeled in that work by the total number of electronic plus photonic indices (${}^1D^{A}_{B}$, ${}^2D^{pA}_{qB}$, ${}^3D^{pqA}_{rsB}$, ${}^3Q^{pqA}_{rsB}$, and ${}^3G^{pqA}_{rsB}$), are the dressed forms of the role strings $1$, $x$, $xx$, $oo$, and $xo$, respectively; in particular, the photon RDM, ${}^0{\bf D}$ (${}^1D^{A}_{B}$ in Ref.~\citenum{DePrince22_053710}), is the dressed form of the role string $1$.
Note that conditions on the purely electronic RDMs need not be imposed separately. Each photon-diagonal block of a PSD dressed matrix is PSD, so their sum, the photon partial trace, which recovers the corresponding electronic RDM, is also PSD.

\subsubsection{Anticommutator conditions}
\label{SEC:QED_ANTICOMMUTATOR}

The T1, T2, and T2$'$ conditions do not generalize in the same way as the Hermitian-square conditions in Table~\ref{TAB:POSITIVITY}. These conditions rely on the cancellation of three-body contributions between $\langle CC^\dagger\rangle$ and $\langle C^\dagger C\rangle$. For a dressed operator of the form $C = \sum_I c_I\, O_I\, Y$, where $Y$ is a fixed photon operator, the two terms carry $YY^\dagger$ and $Y^\dagger Y$, respectively, so the cancellation works only when $YY^\dagger = Y^\dagger Y$ ({\em i.e.}, whenever $Y$ is a normal operator). In the dressing used above, $Y = \sum_A c_A |A^\text{p}\rangle\langle 0^\text{p}|$, which is not normal because
\begin{align}
YY^\dagger = \sum_{AB} c_A c_B^* |A^\text{p}\rangle\langle B^\text{p}|
\end{align}
and
\begin{align}
Y^\dagger Y = \sum_A |c_A|^2\, |0^\text{p}\rangle\langle 0^\text{p}|
\end{align}
As a result, the matrix with elements $\langle \{O_I, O_J^\dagger\}\, |A^\text{p}\rangle\langle B^\text{p}|\rangle$ is not PSD for general $N$-representable states. The choice $Y = |A^\text{p}\rangle\langle A^\text{p}|$ yields photon-diagonal conditions of the form
\begin{align}
\label{EQN:PHOTON_DIAGONAL_T}
\langle \{O_I, O_J^\dagger\}\, |A^\text{p}\rangle\langle A^\text{p}| \rangle \succeq 0, \quad A = 0, \dots, n_\text{ph}
\end{align}
which involve only the electron-photon 1RDM and 2RDM; the photon-diagonal T2$'$ condition follows in the same way. Because the sum over $A$ of Eq.~\ref{EQN:PHOTON_DIAGONAL_T} recovers the corresponding electronic condition, the photon-diagonal conditions are at least as strong as the electronic T1, T2, and T2$'$ conditions. Choosing $Y = |\phi^\text{p}\rangle\langle\phi^\text{p}|$ for other normalized photon states $\phi$ yields additional valid conditions that couple different photon-number states. We do not consider these additional rotated conditions here.

\subsubsection{Photon-dressed constraints}
\label{SEC:QED_CONSTRAINTS}

Because the photon density operators commute with the fermionic operators, any operator identity used in Sec.~\ref{SEC:SYMBOLIC} remains valid when both sides are multiplied by $|A^\text{p}\rangle\langle B^\text{p}|$. Each family of linear constraints (Eq.~\ref{EQN:LINEAR_CONSTRAINT}) therefore has a photon-dressed counterpart,
\begin{equation}
\label{EQN:DRESSED_LINEAR_CONSTRAINT}
\sum_k \sum_{K,L} c^{IJ}_{KL}\, {}^kD^{KA}_{LB} - B^{IA}_{JB} = -s^{IJ}\, {}^0D^{A}_{B}
\end{equation}
in which the scalar terms multiply elements of the photon RDM, rather than contributing to $\mathbf{b}$. The coefficients $c^{IJ}_{KL}$ and scalars $s^{IJ}$ are exactly those that arise in the purely electronic constraints, so the photon-dressed constraints require no additional symbolic derivation. Note that, for the photon-diagonal anticommutator matrices (Eq.~\ref{EQN:PHOTON_DIAGONAL_T}), only the $A = B$ members of each family are generated. Expectation value constraints generalize in the same way, so the trace (Eq.~\ref{EQN:TRACE}) and contraction (Eq.~\ref{EQN:CONTRACTION}) constraints become
\begin{align}
0 &= -\eta_\sigma \eta_\tau\, {}^0D^{A}_{B} + \langle N_\sigma N_\tau\, |A^\text{p}\rangle\langle B^\text{p}| \rangle \\
0 &= -\eta_\sigma \langle O_I O_J^\dagger\, |A^\text{p}\rangle\langle B^\text{p}| \rangle + \langle N_\sigma O_I O_J^\dagger\, |A^\text{p}\rangle\langle B^\text{p}| \rangle
\end{align}
and the spin constraints become
\begin{align}
0 &= -S(S+1)\, {}^0D^{A}_{B} + \langle (S_z + S_z S_z + S_- S_+)\, |A^\text{p}\rangle\langle B^\text{p}| \rangle \\
0 &= \langle O_I S_+\, |A^\text{p}\rangle\langle B^\text{p}| \rangle
\end{align}
These expressions recover the contraction and spin constraints of Ref.~\citenum{DePrince22_053710} (in which the photon RDM is denoted ${}^1D^A_B$), {\em e.g.}, the contraction of the electron-electron-photon RDM to the electron-photon RDM, the spin-dependent $S^2$ constraint, and the maximal spin-projection constraint on the electron-hole-photon RDM.  Finally, together with the photon trace (Eq.~\ref{EQN:PHOTON_TRACE}), the dressed constraints imply their purely electronic counterparts, which are recovered by summing the $A = B$ members of each family.

\subsubsection{Block structure and numerical representation}
\label{SEC:QED_NUMERICAL}

The inclusion of photon labels does not impact the symbolic constraint generation described in Sec.~\ref{SEC:SYMBOLIC}. \pq derives each family of constraints once, from purely fermionic operator strings, and the photon structure is introduced only when the expressions are evaluated numerically. The photon labels $A$ and $B$ are appended to the free (row and column) labels of each constraint family, and the composite row index of a photon-dressed matrix is $(o_s + I)(n_{\rm ph}+1) + A$ ({\em cf.} Sec.~\ref{SEC:NUMERICAL}). Each photon-dressed block therefore has dimension $(n_{\rm ph}+1)\, d_b$, so the number of stored elements grows by a factor of $(n_{\rm ph}+1)^2$. Each photon-diagonal anticommutator matrix is stored as $n_{\rm ph}+1$ separate blocks of dimension $d_b$, one per photon-number state. The photon RDM is a single block of dimension $n_{\rm ph}+1$. The Hermitian symmetry of each block is exploited for the composite row and column labels, so constraints are generated only for $(I, A) \le (J, B)$ within diagonal sector pairs, and only for $A \le B$ in families with no fermionic free labels ({\em e.g.}, the trace and $S^2$ constraints). The photon-dependent contributions to the energy (Eqs.~\ref{EQN:PHOTON_HAM_CONTRIBUTION} and \ref{EQN:BLC_CONTRIBUTION}) are evaluated with the same machinery as the electronic terms.

\section{Applications}

\label{SEC:RESULTS}

\subsection{FeMoco}
The FeMoco system has served as a benchmark in electronic structure and quantum information, presenting one of the most challenging ground-state energy tasks to arise in a biochemical system.\cite{li2019electronic, pb2g-j9cw} Specifically, the 76-orbital, 113-electron $S = 3/2$ ground-state problem posed in Ref.~\citenum{li2019electronic} presents a unique challenge of describing the ground-state energy and dominant spin structures of the metalloenzyme active site. Recent work~\cite{zhai2026classical} compares large-scale DMRG calculations, extrapolated to infinite bond dimension, to an active-space-corrected unrestricted coupled-cluster singles-doubles-triples-quadruples method (UCCSDTQ). An attractive feature of v2RDM methods is that they are variational from below, so they can provide a lower-bound certificate to the true ground-state energy. Prior attempts to use v2RDM for the FeMoco problem focused on smaller systems. In 2018, Montgomery and Mazziotti studied a (54o, 54e) model of FeMoco using CASSCF and v2RDM methods with up to a (30o, 30e) subspace.\cite{montgomery2018strong} Here, using our automatically-generated v2RDM code implementing the PQG approximate $N$-representability conditions, we compute a PQG lower-bound estimate (to within the convergence tolerance of the SDP solver) of the ground-state energy of
the full 76-orbital, 113-electron active space of Ref.~\citenum{li2019electronic}.

Table \ref{tab:femoco} provides total energies for the 76-orbital FeMoco model at the PQG, UDMRG, and UCCSDTQ levels of theory (with the latter two energies taken from Ref.~\citenum{zhai2026classical}). We also provide the energy computed using a spin-free approximation to the G condition and in the absence of the D and Q conditions ({\em i.e.}, where the energy, trace, and contraction constraints are expressed in terms of a spin-free ${}^2{\bf G}$ matrix).\cite{DePrince26_8751,pb2g-j9cw} We find that the PQG energy is 642 mE$_\text{h}$ below the extrapolated DMRG energy from Ref.~\citenum{zhai2026classical}. The gap is much more substantial, approaching 4 E$_\text{h}$, when full 2-positivity is approximated by the spin-free generating algebra. The memory efficiency of our automated v2RDM code enabled the execution of the PQG computation on a single H100 GPU card in 24.56 hours, with an average BPSDP iteration time of 3.8 seconds. The BPSDP algorithm required 23078 iterations to converge the primal--dual energy gap to $10^{-5}$ E$_\text{h}$ and the primal and dual errors to $10^{-5}$. The PQG conditions constituted over 299 million double-precision primal variables; the peak GPU memory required by the BPSDP algorithm was 36.9 GB. 

\begin{table}[H]
\caption{Ground-state energies ($\text{E}_\text{h}$) for the 76-orbital, 113-electron model of FeMoco\cite{li2019electronic} from v2RDM calculations compared against reference values from Ref.~\onlinecite{zhai2026classical}.}
\label{tab:femoco}
\begin{ruledtabular}
\begin{tabular}{lr}
Quantity / Method & Energy ($\text{E}_\text{h}$) \\
% \hline
% Core constant (FCIDUMP)                  & $-21021.21401$ \\
\hline
v2RDM (spin-free G) & $-22144.32500$ \\
v2RDM (PQG)                              & $-22141.05217$ \\
Extrapolated UDMRG$^{a}$       & $-22140.4101\phantom{0}$ \\
Composite UCCSDTQ$^{a,b}$ & $-22140.4043\phantom{0}$ \\
\hline
$\Delta E$ (UDMRG $-$ PQG)               & $0.6420$ \\
$\Delta E$ (UDMRG $-$ spin-free G)    & $3.9149$ \\
\end{tabular}
\end{ruledtabular}

$^a$ From Ref.~\onlinecite{zhai2026classical}.

$^b$ Empirical upper bound with high confidence.

\end{table}

\subsection{H$_{10}$ clusters with 4POS}

We now consider the convergence of the v2RDM energy with increasingly strict ensemble $N$-representability conditions. As a test case, we choose another challenging strong correlation problem where v2RDM has been shown to struggle previously: the two-dimensional H$_{10}$ cluster studied by Stair and Evangelista in Ref.~\citenum{Evangelista20_104108}. Specifically, we consider this cluster at a large inter-atom separation (1.80 \AA), where those authors found that v2RDM with the PQG and PQG+T2 conditions gives energies that are 143 and 38 mE$_\text{h}$ below the full CI energy, respectively. This system was revisited using v2RDM with 3POS conditions in Ref.~\citenum{DePrince21_174110}, where it was found that even 3POS leads to an error of 22 mE$_\text{h}$. Table \ref{TAB:H10} provides total energies for this structure, computed using two-particle (PQG), three-particle (3POS), and four-particle (4POS) conditions, as well as the T2 and T2$'$ partial three-particle conditions. All numbers were generated using the current software and verified against existing literature values (PQG and T2;\cite{Evangelista20_104108} 3POS\cite{DePrince21_174110}). Newly reported numbers include the total energy using the T2$'$ and 4POS conditions. T2$'$ offers no improvement over T2 at the reported precision. 4POS, on the other hand, significantly improves the quality of the energy, differing from the full CI energy by roughly 4 mE$_\text{h}$ (see the footnote in Table~\ref{TAB:H10}). This improvement comes at a significant computational cost, however. Table \ref{TAB:H10} also quantifies the problem size associated with each program via the number of primal variables ($n_\text{primal} = \sum_b d_b^2$), the floating-point scaling of the CG and eigendecomposition steps of the BPSDP algorithm, and the time per BPSDP iteration. The number of primal variables grows from only about 89K for PQG to 5M for T2/T2$'$, to 10M for 3POS, and to nearly 700M for 4POS, which also includes the 3POS conditions. The floating-point scalings also increase dramatically with the particle rank of the constraints; the CG and eigendecomposition steps require $\mathcal{O}(n^8)$ and $\mathcal{O}(n^{12})$ floating-point operations for 4POS, and the time per BPSDP iteration reflects these scalings. Computationally, two-positivity constraints are trivial to enforce for this problem; the PQG calculation required about two minutes on an H100 GPU. The problem size is still manageable up to the 3POS level, where the calculation was converged in roughly 42 minutes on the same hardware. The 4POS calculations, however, are nearly intractable. The problem fits in memory on a single H100 card when we use the ``folded'' representation of the problem (see Eqs.~\ref{EQN:Af} and \ref{EQN:AfWAf}), but the BPSDP iterations are slow (41 s), which, when coupled with the characteristically slow convergence of the BPSDP solver, makes it difficult to fully converge the solution. The 4POS energy reported in Table \ref{TAB:H10} corresponds to the value after 50K BPSDP iterations, when the primal-dual energy gap was $\approx 6\times 10^{-4}$ E$_\text{h}$, and the primal and dual errors were $\approx 4\times 10^{-5}$. Because the solution is not converged, neither the primal nor the dual solution is feasible; the tabulated 4POS energy should thus be regarded as a numerical estimate rather than a lower bound.
On the other hand, all other energy values given in Table \ref{TAB:H10} correspond to primal-dual energy gaps of less than $10^{-5}$ E$_\text{h}$ and primal and dual errors below $10^{-5}$. 

\begin{table}[!htbp]
    \centering
    \caption{Energies (E$_\text{h}$) from v2RDM calculations on the sheet structure of H$_{10}$ from Ref.~\citenum{Evangelista20_104108} with an H--H distance of 1.80 \AA, with energy errors with respect to full CI reference calculations (E$_\text{h}$) in parentheses. The number of primal variables in the program ($n_\text{primal}$), the scaling of the CG and eigendecomposition steps in the BPSDP algorithm, and the time per BPSDP iteration (in s) are also provided. }
    \label{TAB:H10}
    \begin{tabular}{cccccc}
        \hline\hline
        & & & \multicolumn{2}{c}{scaling} & \\
        \cline{4-5}
        Method & Energy (Error) & $n_\text{primal}$ & CG & eig &  time / iter \\
        \hline
        PQG  & -4.9132 ($-0.143$)    & 88500      & $n^4$ & $n^6$ & 0.03 \\
        PQG+T2   & -4.8076 ($-0.038$)    & 4698500    & $n^6$ & $n^9$ & 0.07 \\
        PQG+T2$'$  & -4.8076 ($-0.038$)  & 4756700    & $n^6$ & $n^9$ & 0.08 \\
        3POS & -4.7923 ($-0.022$)    & 10176100   & $n^6$ & $n^9$ & 0.22 \\
        4POS & -4.7742$^a$ ($-0.004$)& 689637500  & $n^8$ & $n^{12}$ & 41.00 \\
        
        \hline\hline
    \end{tabular}
    
    $^a$ Not fully converged. After 50K BPSDP iterations, the primal and dual energies were $-4.7742$ and $-4.7735$ E$_\text{h}$, respectively, and the primal and dual errors were $\approx 4\times 10^{-5}$. The tabulated value is the primal energy, which is a numerical estimate rather than a lower bound.
\end{table}

\subsection{Cavity quantum electrodynamics v2RDM}

\begin{table*}[!htpb]
\centering
\caption{Total energies (E$_\text{h}$) along the BeH$_2$ insertion path, forward and reverse barriers along the path ($\Delta E_\text{f}$ and $\Delta E_\text{r}$, mE$_\text{h}$), and cavity-induced changes to the barriers ($\Delta\Delta E_\text{f}$ and $\Delta\Delta E_\text{r}$, mE$_\text{h}$). The symbol $n_\text{ph}$ denotes the maximum photon number; $n_\text{ph} = 0$ is the cavity-free case. All calculations use the STO-3G basis set. For calculations involving a cavity, the mode is polarized along the $x$-direction, with $\omega = 2$ eV and $\lambda = 0.05$ a.u.}
\label{TAB:BEH2_X}
\begin{tabular}{lccccccccc}
\hline
\hline
 & \multicolumn{4}{c}{$n_\text{ph}$ = 0} &~&  \multicolumn{4}{c}{$n_\text{ph}$ = 1} \\
 \cline{2-5} \cline{7-10}
$x$                       &   PQG     &  PQG+T2        & 3POS      & full CI       &~& PQG       & PQG+T2        & 3POS      & full CI       \\
\hline
0.0                       & -15.596038 & -15.594743  & -15.594722 & -15.594721 &~& -15.590868 & -15.589658 & -15.589635   & -15.589635 \\ 
1.5                       & -15.390553 & -15.374856  & -15.373248 & -15.373246 &~& -15.384344 & -15.368548 & -15.366762   & -15.366760 \\
2.117                     & -15.513723   & -15.512816 & -15.512799 & -15.512798 &~& -15.508458 & -15.507709 & -15.507695   & -15.507695 \\
$\Delta E_\text{f}$       &     123.2 &     138.0  &     139.6 &     139.6 &~&     124.1 &       139.2 &     140.9 &     140.9 \\
$\Delta E_\text{r}$       &     205.5 &     219.9  &     221.5 &     221.5 &~&     206.5 &       221.1 &     222.9 &     222.9 \\
$\Delta\Delta E_\text{f}$ &         - &         -  &         - &         - &~&       0.9 &       1.2 &       1.4   &       1.4 \\
$\Delta\Delta E_\text{r}$ &         - &         -  &         - &         - &~&       1.0 &       1.2 &       1.4   &       1.4 \\
\hline
\hline
\end{tabular}
\end{table*}

Let us now consider the generalization of the v2RDM approach for cavity QED problems using the photon-dressed $N$-representability constraints described in Sec.~\ref{SEC:QED}. Reference \citenum{DePrince22_053710} applied the photon-dressed PQG conditions to two well-known strong correlation problems: the symmetric dissociation of linear hydrogen chains and the singlet-triplet gap in the linear oligoacene series. Here, we apply the photon-dressed PQG, T2, and 3POS constraints to another classic model for benchmarking multireference correlation methods: the C$_{2v}$ insertion pathway for BeH$_2$.\cite{Bartlett83_835} We use a single reaction coordinate, $x$, to characterize the pathway, with the coordinates for the atoms defined (in units of~\AA) as
\begin{eqnarray}
&\text { Be:}& \text{(0.0, 0.0, 0.0)}, \nonumber \\
&\text { H: }& \text{($x$,  1.344 - 0.46$x$, 0.0)}, \nonumber \\
&\text { H: }& \text{($x$, -1.344 + 0.46$x$, 0.0)}.
\end{eqnarray}
Table \ref{TAB:BEH2_X} provides energies at three points along this pathway. At $x = 0.0$~\AA, we have the linear BeH$_2$ molecule; $x = 1.5$~\AA\ corresponds to the transition state along the pathway; and, at $x = 2.117$~\AA\ (4 $a_0$), the endpoint of the model pathway, we have H$_2$ at its equilibrium bond length (0.74~\AA) as it approaches the Be atom. In Table \ref{TAB:BEH2_X}, the symbol $\Delta E_\text{f}$ represents the forward reaction barrier (from $x = 2.117$~\AA\ to $x = 1.5$~\AA), whereas $\Delta E_\text{r}$ denotes the reverse barrier (from $x = 0.0$~\AA\ to $x = 1.5$~\AA). For calculations involving the cavity, the photon mode is polarized along the $x$-axis and has a frequency $\omega = 2$ eV. The cavity coupling strength is $\lambda = |{\bm \lambda}| = 0.05$ a.u., which corresponds to a ``picocavity''\cite{Baumberg22_5859} setup. We consider the singlet states along the reaction pathway, with appropriate photon-dressed spin constraints included in the SDP. All calculations used the STO-3G basis set, and the photon mode was represented in a photon-number basis truncated at $n_\text{ph} = 1$. Most of the data in this table come from calculations that were considered converged when the primal--dual energy gap fell below $10^{-5}$ E$_\text{h}$ and the primal and dual errors fell below $10^{-5}$. However, for calculations involving the PQG and T2 conditions at $x = 1.5$~\AA, the energy gap threshold was loosened to $5\times 10^{-5}$ E$_\text{h}$. Table \ref{TAB:BEH2_X} also includes QED full CI\cite{Foley24_1214,Foley25_8812} data obtained using the \texttt{qed-ci} package.\cite{FoleyLab_qedci}

In the absence of the cavity, the PQG conditions, as expected, substantially overestimate electron correlation at all considered geometries. Specifically, PQG energies are below the full CI ones by 1.3, 17.3, and 0.9 mE$_\text{h}$ at $x = 0.0$, 1.5, and 2.117~\AA. The error is largest at the transition state, so PQG underestimates the forward and reverse barriers by 16.4 and 16.0 mE$_\text{h}$, respectively. The T2 condition dramatically reduces the absolute energy errors to 0.02, 1.6, and 0.02 mE$_\text{h}$ at these geometries, and the barriers are predicted to within 1.6 mE$_\text{h}$ of the full CI values. 3POS is essentially exact; the corresponding energies agree with those from full CI to within 2 $\mu$E$_\text{h}$, which is well within the convergence threshold of the SDP solver. 

The energies at all geometries and from all levels of theory increase when the molecule is coupled to the cavity mode. The cavity effect is largest at the transition state, so the forward and reverse barriers increase slightly at $\lambda = 0.05$ a.u. At the full CI level, the cavity increases the forward and reverse barriers by 1.4 mE$_\text{h}$. Energies from photon-dressed v2RDM calculations show the same trends as in the cavity-free case. PQG energies are 1.2, 17.6, and 0.8 mE$_\text{h}$ lower than those from full CI at $x = 0.0$, 1.5, and 2.117~\AA, and PQG underestimates the cavity effect on the barrier heights, predicting changes of only 0.9 and 1.0 mE$_\text{h}$ for the forward and reverse barriers. At the PQG+T2 level, we find that the cavity-induced changes to both barriers are 1.2 mE$_\text{h}$, which is in much better agreement with the full CI. Again, 3POS is essentially exact, in terms of both the total energies and cavity-induced changes to the barriers. Lastly, we note that, even at $n_\text{ph} = 1$, the calculations are already nearly converged with respect to the size of the photon basis.  Increasing the photon-number basis to $n_\text{ph} = 2$ and 3 changes the photon-dressed PQG energies by less than $10^{-5}$ E$_\text{h}$. The full CI energies change by at most $5\times 10^{-5}$ E$_\text{h}$ under the same changes to the photon basis, and the corresponding cavity-induced modifications to the barriers change by less than 0.05 mE$_\text{h}$.

\section{Conclusions}

\label{SEC:CONCLUSIONS}

We have presented an automated framework for generating semidefinite programs that arise in v2RDM theory using the \pq second-quantized algebra engine. Each positivity condition is specified by a string of creation and annihilation operators, from which \pq derives the linear maps to the all-particle RDMs, with appropriate spin-block structures. Trace, contraction, and spin-symmetry constraints are implemented using the same machinery. These expressions are then assembled into a sparse-matrix representation of the SDP and solved with a boundary-point SDP algorithm that we have implemented in the \texttt{libsdp} library. The machinery we have developed extends easily to mixed fermion--boson systems, where we have developed photon-dressed analogs of the standard electronic $N$-representability conditions. Because each new condition requires only the specification of its generating operators, \pq enables the rapid design and implementation of new $N$-representability conditions, without the tedious and error-prone derivation of the underlying linear maps. When coupled to the GPU-accelerated BPSDP solver in \texttt{libsdp}, the resulting SDPs can also be solved efficiently at scale; for example, the PQG calculation on the 76-orbital model of FeMoco, which involves roughly $3\times 10^{8}$ primal variables, requires about 25 hours on a single H100 GPU. Both \pq and \texttt{libsdp} are  available on GitHub,\cite{deprince_pdaggerq,deprince_libsdp} and the \pq repository includes example scripts that generate v2RDM SDPs and solve them with the BPSDP solver in \texttt{libsdp}.

\vspace{0.5cm}

\noindent {\bf Acknowledgments.}
The cavity QED aspects of this work are based upon work supported by the National Science Foundation under Grant No.~CHE-2533603.

\vspace{0.5cm}

\noindent {\bf AI Statement.} 
Generative AI tools were used in developing the software described in this work. Google DeepMind's Gemini (version 3.1 Pro, accessed July--August 2026) was used as a programming assistant in early versions of the SDP generator and the JAX/GPU implementation of the BPSDP solver, which were subsequently refactored using Claude Opus 5.5 (Anthropic, accessed through the Claude Code interface) to implement the role-string-based SDP generator and its cavity QED extension, as well as to optimize the memory usage and performance of \pq and the BPSDP solver. These tools were used to accelerate code development. All AI-generated code was reviewed and revised by the authors and validated against independent (QED-)v2RDM implementations developed by the authors and their collaborators. The authors take full responsibility for the results reported here.

\vspace{0.5cm}

\noindent {\bf Data Availability.} The data that support the findings of this study are available from the corresponding author upon reasonable request.

\bibliography{main.bib}

\end{document}